\documentclass[twocolumn,prd]{revtex4}
\usepackage{graphicx}
\newcommand{\beq}{\begin{equation}}
\newcommand{\eeq}{\end{equation}}
\newcommand{\beqa}{\begin{eqnarray}}
\newcommand{\eeqa}{\end{eqnarray}}

\def\l({\left[}
\def\r){\right]}
\def\ltap{\raisebox{-.55ex}{\rlap{$\sim$}} \raisebox{.4ex}{$<$}}
\def\gtap{\raisebox{-.55ex}{\rlap{$\sim$}} \raisebox{.4ex}{$>$}}
\def\gsim{\mathrel{\gtap}}
\def\lsim{\mathrel{\ltap}}

\begin{document}

\title{TeV signatures of compact UHECR accelerators}
\author{A.~Neronov$^1$, P.~Tinyakov$^{2,4}$ and I.~Tkachev$^{3,4}$}
\affiliation{$^1$Institute of Theoretical Physics, EPFL, BSP, CH-1015
Lausanne, Switzerland\\ $^2$ Service de Physique Th\'eorique, CP 225,
Universit\'e Libre de Bruxelles, B-1050, Brussels, Belgium\\ $^3$ 
Department of Physics, Theory Division, CERN, 1211 Geneva 23,
Switzerland\\ $^4$
Institute for Nuclear Research of the Russian Academy of Sciences, 117312,
Moscow, Russia }
\begin{abstract}
We study numerically particle acceleration by the electric field induced near
the horizon of a rotating supermassive ($M\sim 10^9-10^{10}M_{\odot}$) black
hole embedded in the magnetic field $B$. We find that acceleration of protons
to energy $E\simeq 10^{20}$~eV is possible only at extreme values of $M$ and
$B$.  We also find that the acceleration is very inefficient and is
accompanied by a broad band MeV-TeV radiation whose total power exceeds at
least by a factor of 1000 the total power emitted in ultra-high energy cosmic
rays (UHECR). This implies that if $O(10)$ nearby quasar remnants were sources
of proton events with energy $E>10^{20}$~eV, then each quasar remnant would
overshine e.g. the Crab nebula by more than two orders of magnitude in the TeV
energy band. Recent TeV observations exclude this possibility. A model in
which $O(100)$ sources are situated at $100-1000$~Mpc is not ruled out and can
be experimentally tested by present TeV $\gamma$-ray telescopes. Such a model
can explain the observed UHECR flux at moderate energies $E\sim (4-5)\times
10^{19}$~eV.
\end{abstract}

\maketitle

\section{Introduction}

A conventional hypothesis of ultra-high energy cosmic ray (UHECR) acceleration
in extragalactic astrophysical objects has two important consequences. First,
it predicts the Greisen-Zatsepin-Kuzmin (GZK) cutoff \cite{GZK} in the
spectrum of UHECR at energy of order $5\times 10^{19}$~eV.  Whether such a
cutoff indeed exists in nature is currently an open question
\cite{AGASA,hires}. Second, it implies that the observed highest-energy cosmic
rays with $E>10^{20}$~eV should come from within the GZK distance $\sim
50$~Mpc. Moreover, under plausible assumptions about extragalactic magnetic
fields supported by recent simulations \cite{IGMF}, the propagation of UHE
protons over the GZK distance is rectilinear and the observed events should
point back to their sources. While sub-GZK UHECR were found to correlate with
BL Lacertae objects \cite{bllacs,EgretBllacs}, no significant correlations of
cosmic rays with energies $E \gsim 10^{20}$ eV with nearby sources were found
\cite{AGASA2}.

In view of the latter problem, a question arises whether there exist UHECR
accelerators which can produce super-GZK protons and are quiet in the
electromagnetic (EM) channel. If such quiet accelerators existed, they could
explain the apparent absence of sources within $\sim 50$~Mpc in the direction
of the highest-energy events. This idea was advocated, e.g., in
Ref.~\cite{dead} where sources of UHE protons were associated with
supermassive black holes in quiet galactic nuclei (the so-called ``dead
quasars'').  However, it was pointed out \cite{Levinson:nx} that most of the
energy available for particle acceleration in such an environment is spent for
EM radiation by the accelerated particles. As a consequence, the flux of TeV
$\gamma$-rays produced by  such an accelerator may be at a detectable
level.

Recent observations by HEGRA/AIROBICC \cite{hegralimit}, MILAGRO
\cite{milagro} and TIBET \cite{tibet} arrays improved substantially upper
limits on the flux of $\gamma$-rays above $10$ TeV from the point sources in
the Northern hemisphere. This may exclude completely a possibility to explain
the observed super-GZK cosmic rays by the acceleration near supermassive black
holes. The purpose of this paper is to analyze this question
quantitatively. To this end we study particle acceleration near the black hole
horizon numerically. Following Refs. \cite{dead,Levinson:nx} we restrict
ourselves to the case of protons. The case of heavy nuclei acceleration,
propagation and detection is phenomenologically very different and requires
separate consideration \cite{longpaper}. In particular, heavy nuclei can be
easily desintegrated already at the stage of acceleration.  

We would like to stress that it is not our pupose to construct a realistic
model of a compact UHE proton accelerator, but rather to find out whether
quiet compact accelerators may exist, even if most favorable conditions for
acceleration are provided. To this end we minimize the energy losses of
accelerated particles by considering acceleration in ordered electromagnetic
field and neglect all possible losses related to scattering of the
accelerated particles off matter and radiation present in the acceleration
site. However, we take into account, in a self-consistent way, the
synchrotron/curvature radiation losses which are intrinsic to the acceleration
process.  Clearly, this approximation corresponds to most favorable conditions
for particle acceleration. In realistic models the resulting particle energy
will be smaller, while emitted EM power larger. Therefore, our results should
be considered as a lower bound on the ratio of electromagnetic to UHECR power
of a cosmic ray accelerator based on a rotating supermassive black hole.

We find that the flux produced by a nearby UHE proton
accelerator of super-GZK cosmic rays in the energy band $E_{\gamma}>10$~TeV
should be at least $100-1000$ times larger than that of the Crab nebula. The
existence of such sources is indeed excluded by recent observations
\cite{hegralimit,milagro}.  At the same time, the constraints on the sources
of sub-GZK cosmic rays are weaker or absent, see Sect.~VII for details.

The paper is organized as follows. In Sect.~II we describe our minimum loss
model in more detail. In Sect.~III we present the analytical estimate and the
numerical calculation of maximum particle energy. In Sect.~IV the
self-consistency constraints on the parameters of this model are considered
which arise from the requirement that there is no on-site $e^+e^-$ pair
production caused by emitted radiation.  In Sect.~V the calculation of the EM
luminosity of the accelerator is presented.  In Sect.~VI observational
constraints are derived. Sect.~ VII contains the discussion of the results and
concluding remarks.

\section{The model}

The model we consider is based on a rotating supermassive black hole embedded
in a uniform magnetic field. Because of the rotational drag of magnetic field
lines, the electric field is generated and leads to acceleration of
particles. In the absence of matter, the corresponding solution to the
Einstein-Maxwell equations is known analytically at arbitrary inclination
angle of black hole rotation axis with respect to the magnetic field
\cite{field-config,field-config2}. We assume low accretion rate and small
matter and radiation density near the black hole, and neglect their back
reaction on the EM and gravitational fields.  We also neglect the effect of
matter on propagation of accelerated protons. This corresponds to most
favorable conditions for particle acceleration, and therefore leads to a
maximum proton energy and minimum EM radiation.

The model has three parameters: the black hole mass $M$, the strength of the
magnetic field $B$ and the inclination angle $\chi$. We consider the maximally
rotating black hole with rotation moment per unit mass $a=M$. This maximizes
the strength of the rotation-induced electric field.  For a given injection
rate and geometry, the above parameters completely determine the trajectories
of accelerated particles and, therefore, their final energies and the emitted
radiation. We reconstruct particle trajectories numerically keeping track of
emitted radiation and taking into account its back reaction onto particle
propagation.

We assume that protons flow into the acceleration volume from the accretion
disk which is situated at larger radii.  We model this accretion by injecting
non-relativistic particles uniformly over the sphere of the Schwarzschild
radius $R_{\rm S} =2GM$, which is two times larger than the horizon of the
maximally rotating black hole.  We follow the trajectories of particles which
propagate {\it toward} the horizon and then are expelled from the vicinity of
the black hole with high energies. It turns out that such trajectories exist
only if the inclination angle of magnetic field with respect to rotation axis
is large enough, $\chi\gsim 10^{\circ}$.  For smaller inclination angles all
particles which propagate toward the horizon are finally absorbed by the black
hole. This means that the stationary regime in which particles accreted onto
the black hole are subsequently accelerated and ejected with high energies
exists only at $\chi\gsim 10^{\circ}$.  In this regime the change of
inclination angle and injection radius do not affect strongly the maximum
energies of particles. The details of the particle acceleration and the
description of the numerical procedures will be given elsewhere
\cite{longpaper}.

\section{Maximum particle energies in the stationary regime}
\label{sect:losses}

In the abscence of matter and radiation backgrounds, particle energies are
limited by the radiation loss intrinsic to the acceleration process.  For a
general electric and magnetic field configuration the energy loss in
the ultra-relativistic limit is given by
\cite{landau}
\begin{equation}
\label{eq:loss_gen}
\frac{d{\cal E}}{dt}=-\frac{2e^4{\cal E}^2}{3m^4} \left[ (\vec E+\vec
v\times\vec B)^2-(\vec E\vec v)^2\right],
\end{equation}
where $m$, $e$ and $\vec v$ are particle mass, charge and velocity,
respectively. We use this equation in our numerical modelling to calculate the
electromagnetic radiation produced by the accelerated particles and to account
for the back reaction of this radiation on particle trajectories.

Before presenting the numerical results it is useful to summarize some simple
qualitative estimates, see e.g. \cite{Hillas:is,Levinson:nx,Aharonian:2002we}.
Consider particle acceleration by a generic electromagnetic field obeying
$|\vec E|\sim |\vec B|$.  If the energy losses can be neglected, energies of
accelerated particles are estimated as
\begin{equation}
\label{potential}
{\cal E}= eBR\simeq 10^{22} \left[{B\over 10^4\, \mbox{G}}\right]
\left[\frac{M}{10^{10}M_\odot}\right]\mbox{ eV.}
\end{equation}
where we have assumed that the size of acceleration region $R$ is of
order of the gravitational radius of the black hole $R\simeq 2GM$.
However, if the magnetic field strength is high, one cannot neglect
the synchrotron/curvature energy losses. 

 If no special relative orientation of the three vectors
$\vec E$, $\vec B$ and $\vec v$ is assumed,
Eq. (\ref{eq:loss_gen}) becomes 
\begin{equation}
\frac{d{\cal E}}{dt}\simeq -\frac{2e^4B^2{\cal E}^2}{3m^4} ,
\label{eq:synch}
\end{equation}
which is the standard formula for the synchrotron energy loss.  Equating the
rate of energy gain $d{\cal E}/dt=eE\sim eB$ to the rate of energy loss, one
finds that in the synchrotron-loss-saturated regime the maximum energy is
given by (see e.g. \cite{medvedev})
\begin{equation}
\label{eq:synch_cut}
{\cal E}_{\rm syn}=\left[\frac{3m^4}{2e^3B_0}\right]^{1/2}
\simeq 1.6\times 10^{18} \, 
\left[{B\over 10^4\mbox{ G}}
\right]^{-1/2}\mbox{ eV.}
\end{equation}
Here we have assumed that accelerated particles are protons; for
electrons the maximum energy is much smaller.

The critical magnetic field strength at which the synchrotron energy loss
becomes important can be found from the condition that the estimates 
(\ref{potential}) and (\ref{eq:synch_cut}) give 
the same result,
\begin{equation}
B_{\rm crit} = \left[{3m^4 \over 2e^5R^2}\right]^{1/3} 
\simeq 30\left[{M\over 10^{10}M_{\odot}}\right]^{-2/3} \;{\rm G}.
\label{eq:Bcrit}
\end{equation}
Here it is assumed that $R\sim R_{\rm S}$. 
This critical field corresponds to particle energy 
\begin{equation}
\label{ecrit}
{\cal E}_{\rm crit}\approx 3\times 10^{19}\left[{M\over 10^{10}M_{\odot}}
\right]^{1/3}
\mbox{ eV}, 
\end{equation} 
which is a maximum energy attainable in the synch\-ro\-tron-loss-saturated
regime for a given black hole mass.

Acceleration is more efficient (the loss (\ref{eq:loss_gen}) can be orders
of magnitude smaller) in the special
case when $\vec E$, $\vec B$ and  $\vec v$ are nearly aligned. 
These conditions may be approximately
satisfied in some regions around the black hole. In this case
particles follow closely the curved field lines and curvature
radiaiton loss 
\begin{equation}
\frac{d{\cal E}}{dt}=-\frac{2e^2{\cal E}^4}{3m^4R^2},
\end{equation}
becomes the main energy loss channel for high-energy particles. For an
order-of magnitude estimate one can assume that the curvature scale of
the magnetic field lines is of the order of the size of acceleration
region, $R\simeq 2GM$.  This translates into the following maximum
energy,
\begin{equation}
{\cal E}_{\rm cur}=\left[{3m^4 R^2 B \over 2e}\right]^{1/4},
\label{eq:Emaxcur}
\end{equation}
which gives for protons
\begin{equation}
\label{curv_cut}
{\cal E}_{\rm cur}=1.1\times
10^{20}\left[\frac{M}{10^{10}M_\odot}\right]^{1/2}
\left[\frac{B}{10^4\mbox{ G}}\right]^{1/4}\mbox{ eV}.
\end{equation}
where we have set again $R= R_{\rm S}$.  The
range of applicability of eqs.(\ref{eq:Emaxcur}) and (\ref{curv_cut}) is
given by the same condition $B>B_{\rm crit}$.
 
In the numerical simulations, we injected protons uniformly over the sphere
surrounding the black hole. We disregarded trajectories which start at
injection sphere and move outward.  Among protons which approach the horizon
and are then expelled to infinity, we selected those which have maximum final
energy. For the black hole mass $M=10^{10}M_{\odot}$, the dependence of this
maximum energy on the strength of the magnetic field is shown in Fig.~1 by red
line (upper curve). For energies of order $10^{20}$~eV and higher, the
numerically calculated curve approaches the limit (\ref{curv_cut}) which
corresponds to the curvature-loss-saturated regime. The acceleration to these
energies requires magnetic fields in excess of $10^4$~G. The necessary
magnetic field is even stronger for smaller black hole masses,
cf. Eq.(\ref{curv_cut}). The maximum energies of protons do not depend
strongly on the inclination angle in a wide range of $\chi$.
\begin{figure}
\resizebox{\hsize}{!}{\includegraphics[angle=0]{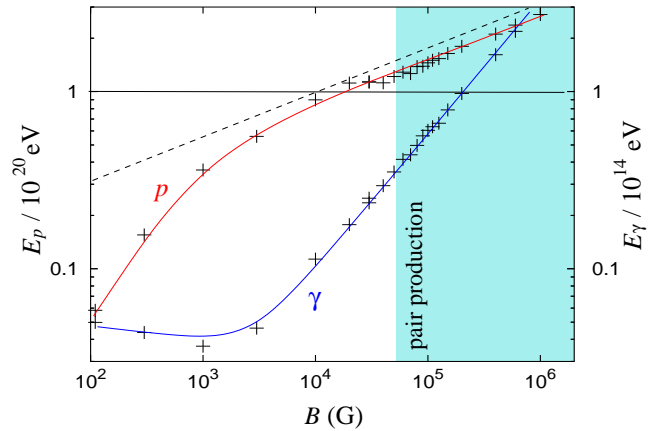}}
\caption{The results of numerical calculation of maximum energies of
accelerated protons and accompanying $\gamma$-rays are shown by crosses (solid
lines are fits to numerical data). The dashed line is the estimate
(\ref{curv_cut}) of the proton energies in the curvature-loss dominated
regime. The shaded region corresponds to the magnetic field strength exceeding
the threshold of pair production by the curvature $\gamma$-rays. The black
hole mass is $M=10^{10}M_\odot$.}
\label{fig:energy}
\end{figure}

\section{Constraints from pair production}
\label{sect:discharge}

There is an important self-consistency constraint which does not allow
to increase $B$ and $M$ independently in order to reach higher
energies. The reason is as follows. In our model it was assumed
that the acceleration proceeds in the vacuum.  However, at a
sufficiently strong magnetic field, photons of curvature radiation may
produce $e^+e^-$ pairs. Electrons and positrons will be accelerated in
turn and produce more photons, which will again produce $e^+e^-$
pairs, etc. The plasma created by this cascade will then neutralize
the electric field and prevent further acceleration of particles.
For consistency of the model we have to require that the cascade does
not develop. 

Consider this process in more detail. The energy $\epsilon_{\gamma}$
of the curvature photons in the regime when particle energies are
limited by curvature losses is estimated as
\[
\epsilon_{\gamma}= 
{3 {\cal E}_{\rm cur}^3\over 2 m^3R}\propto B^{3/4} M^{1/2}. 
\]
(cf. Eq.~(\ref{eq:Emaxcur})).
Remarkably, the photon energy does not depend on the particle mass. This
means that proton-originated and electron-originated photons have the
same energy. Numerically, one has
\begin{equation}
\epsilon_{\gamma} \simeq 14  \left[{B\over 10^4\, {\rm G}} 
\right]^{3/4}
\left[{M\over  10^{10}M_{\odot}}\right]^{1/2} \mbox{TeV}\; .
\label{eq:photon-energy}
\end{equation}
If this energy is enough to produce more than one $e^+e^-$ pair
within the acceleration site, the instability may develop.

Therefore, for the stationary operating accelerator the mean free path 
$d$ of a $\gamma$-ray in the background
of a strong magnetic field  (see \cite{erber}) has to be larger than the size
of the acceleration region 
\beq
\label{dgamma}
d\approx 100 \left[{10^4\, {\rm G}\over B}\right] 
\exp\left(8m_e^3\over 3eB\epsilon_\gamma\right)
\, \mbox{ cm}> R_{\rm S}. 
\eeq
This requirement leads to the condition
\begin{equation}
B < 3.6\times 10^4\left[\frac{10^{10}M_\odot}{M}\right]^{2/7}\mbox{ G.}
\label{eq:discharge_cond}
\end{equation}
on the magnetic field in the vicinity of the horizon.

In numerical calculation of proton trajectories, we kept track of the emitted
photons. For given parameters of the accelerator, we determined the maximum
photon energy.  The dependence of this energy on the magnetic field strength
is shown in Fig.~\ref{fig:energy} by green line (lower curve).  Substituting
the calculated photon energy into Eq.~(\ref{dgamma}) we can check whether the
accelerator is in the stationary regime. The shaded region in
Fig.~\ref{fig:energy} corresponds to non-negligible pair production. The
results of numerical calculation are in good agreement with
Eq.~(\ref{eq:discharge_cond}).

From Fig.~\ref{fig:energy} we conclude that acceleration of protons to
energies higher than $10^{20}$~eV is marginally possible in a small region of
parameter space $(M,B)$. The magnetic field strength $B$ should be close to the
threshold of pair production. The black hole mass $M$ should be larger than
$10^{10}M_\odot$. Such black holes are rare. For example, in Ref
\cite{BHmasses} it is found that supermassive black holes in AGNs range in
$10^{6.5}-10^{10.2}M_\odot$ with the mean mass being $10^{8.9}M_\odot$.  The
list of nearby (within 40~Mpc) candidates for quasar remnants
\cite{Torres:2002bb} does not contain black holes with masses above $5\times
10^8M_\odot$.
  
Under reasonable assumptions about the black hole mass, the acceleration to
energies above $10^{20}$ eV is not possible in the stationary regime discussed
above (no particle production in acceleration volume).  However, in spite of
the fact that the accelerator can not operate permanently, it is possible that
UHECR are produced during ``flares'', or short episodes of activity of the
accelerator which are interrupted by discharges. The natural duration of one
flare is about the time needed for the charge redistribution 
and neutralization of electric field in the acceleration volume to
establish. This
can be roughly estimated as the light-crossing time $T_{\rm flare}\sim R_{\rm
S}/c\approx 10$ hours for the $3\times 10^{9}M_\odot$ black hole. During such
flares the electromagnetic luminosity of the accelerator would be
much higher than the luminosity produced in the stationary regime, because the
electromagnetic flux is dominated by the radiation produced by $e^+e^-$ pairs
whose number density is much higher than the density of initial protons. Since
we are interested in the possibility of having a ``quiet'' UHE proton
accelerator,
we concentrate in the next Section on the case of stationary regime when the
parameteres of the model are tuned to $B\approx 3\times 10^4$~G, $M\approx
10^{10}M_\odot$.

\section{Electromagnetic luminosity of the accelerator}
\label{EM}

It is clear from Fig.~\ref{fig:energy} that the acceleration of protons to
energies above $10^{20}$~eV proceeds in the curvature-loss-saturated
regime. In this regime most of the work done by the accelerating electric
field is spent on the emission of curvature readiation rather than on the
increase of particle energy.  The ratio of the dissipated energy to the final
energy of proton is
\begin{equation}
{\cal R}
\simeq
{eBR\over {\cal E}_{\rm cur}} \simeq
2\times 10^2 \left[ {M\over 10^{10}M_{\odot}}\right]^{1/2}
\left[ {B\over 3\times 10^4\, {\rm G} }\right]^{3/4}.
\label{eq:E_EM/E_UHECR}
\end{equation}
Thus, the energy carried away by photons is at least hundred times higher than
the energy carried by cosmic rays. Since only a small fraction of the
accelerated protons reaches the UHECR energies ${\cal E}\ge 10^{20}$ eV, the
ratio of the electromagnetic luminosity of the accelerator to its 
luminosity in UHECR with $E>10^{20}$~eV is even higher.

Numerically, we calculated this ratio as follows.  We summed energies of those
protons which were accelerated above $10^{20}$~eV, and summed energy emitted
in synchrotron/curvature radiaiton (including the radiation 
emitted by protons which
did not acquire sufficient energy while being expelled to infinity). We then
took the ratio of the two sums.

\begin{figure}
\resizebox{\hsize}{!}{\includegraphics[angle=0]{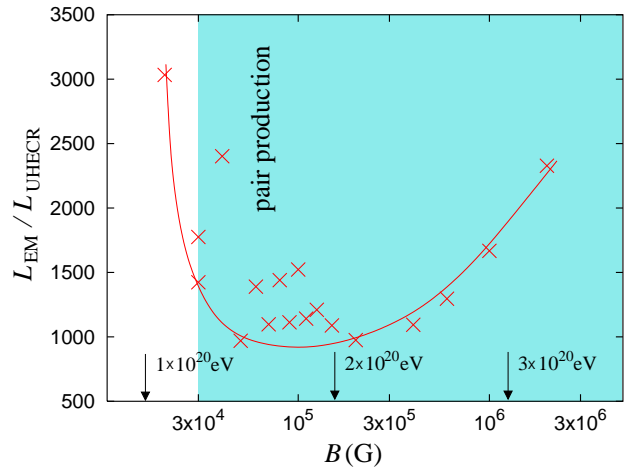}}
\caption{Numerically calculated ratio of electromagnetic luminosity of the
accelerator to the luminosity emitted in particles with energies ${\cal E}\ge
10^{20}$~eV is shown by crosses, with the solid line being the fit to the
numerical results. The shaded region corresponds to the magnetic field
strength above the threshold of pair production by the curvature
$\gamma$-rays. The black hole mass is $M=10^{10}M_\odot$.}
\label{fig:ratio}
\end{figure}

The results of numerical calculation of the ratio of electromagnetic and UHECR
luminosities in the stationary regime are shown in Fig.~\ref{fig:ratio} by
crosses.
Variations are due to fluctuations in the precise positions of the injection
points \footnote{For this calculation we performed injection in $10^3$
randomly chosen points uniformly distributed over the sphere.}.  The
qualitative behavior of the numerical results is easy to understand. Close to
the ``threshold'' ${\cal E}=10^{20}$~eV, the accelerator emits finite power
$L_{\rm EM}$ but does not produce UHECR with ${\cal E}>10^{20}$ eV. Therefore,
the ratio $L_{\rm EM}/L_{UHECR}$ diverges at ${\cal E}\rightarrow 10^{20}$~eV.
If the magnetic field is large, the maximum energies of particles increase as
well. However, the ratio ${\cal R}$ also increases for each particle according
to Eq.~(\ref{eq:E_EM/E_UHECR}), and so does $L_{\rm EM}/L_{UHECR}$.  The
minimal value of $L_{\rm EM}/L_{\rm UHECR}$ is reached at ${\cal E}\approx
1.5\times 10^{20}$~eV. The numerically calculated minimum of $L_{\rm
EM}/L_{\rm UHECR}$ is by a factor of 10 larger than the estimate
(\ref{eq:E_EM/E_UHECR}).

When obtaining the results of Fig.~\ref{fig:ratio} we have taken into account
only the curvature radiation produced by protons. For the magnetic field
strength above the threshold of pair production $\sim 3 \times 10^4$~G (and,
correspondingly, ${\cal E}_{\rm max}>1.3\times 10^{20}$ eV, see
Fig. \ref{fig:energy}), our results give the {\em lower bound} on the
electromagnetic luminosity of the compact accelerator.

One can see from Fig.~\ref{fig:ratio} that the electromagnetic luminosity of
UHE proton accelerator based on a rotating supermassive black hole is
\begin{equation}
L_{\rm EM} \gsim 10^3\, L_{\rm UHECR} \; .
\label{eq:TeV_luminosity}
\end{equation}
Since the typical energy of photons of curvature radiation is about
10~TeV (see Eq.~(\ref{eq:photon-energy})), the above relation implies
that such a source of UHE protons would be much more powerful in the 10 TeV
band than in the UHECR channel.

\section{Observational constraints}
\label{constraints}

The fact that production of UHE protons in the ``quiet'' accelerator is
accompanied by emission of the TeV $\gamma$-ray flux enables us to put strong
constraints on the possibility of existence of such accelerators in the nearby
Universe.  Following Ref.~\cite{dead}, let us assume that there are about
$N\sim 10$ nearby UHECR accelerators which are not further than $D_{\rm
GZK}\sim 50$~Mpc from the Earth. If these sources give the major contribution
to the flux of cosmic rays above $10^{20}$~eV \cite{uhecr_flux},
\[ 
F_{\rm UHECR}^{tot}\sim
(2\div 6)\times 10^{-11}\; {\rm \frac{erg}{s~cm^{2}}},
\]
the mean energy flux produced by each source is
\begin{equation}
\label{mean}
F_{\rm UHECR} \sim {F^{tot}_{\rm UHECR} \over N}
\sim (2\div 6)\times 10^{-12}\; {\rm \frac{erg}{s~cm^{2}}}.
\end{equation} 
We have seen in the previous section that the lower bound on the
ratio of electromagnetic and UHECR luminosity of such a source is ${\cal
R}>10^{3}$. This means that the electromagnetic flux from the source should be
\begin{equation}
\label{tev}
F_{\rm EM} > (2\div 6)\times 10^{-9}\, \left[ {10\over N}\right]
 \;{\rm \frac{erg}{s~cm^{2}}}, 
\end{equation}
which implies the total luminosity larger than 
\begin{equation}
L_{EM}\sim 10^{44} \mbox{erg/s}
\end{equation}
at the distance $D_{\rm GZK}$.  This would be an extremely powerful source of
TeV radiation.  For comparison, the flux of the Crab nebula at energies above
$15$~TeV is $F_{\rm Crab} \sim 10^{-11}$~erg~cm$^{-2}$~s$^{-1}$
\cite{hegralimit}. Thus, the hypothetical ``quiet'' cosmic ray sources which
would explain the observed UHECR flux should be 100-1000 times brighter in the
TeV band than the Crab nebula.

The possibility of existence of persistent point sources of this type in the
Northern hemisphere is excluded by the measurements of HEGRA AIROBICC Array
\cite{hegralimit} and by MILAGRO experiment \cite{milagro,milagro1}. The upper
limit on the energy flux from an undetected point source of $\sim 15$~TeV
$\gamma$-rays provided by HEGRA/AIROBICC group \cite{hegralimit} is at the
level of $F_{HEGRA}\lsim (2\div 3)\, F_{\rm Crab}$. Much tighter upper limit
was published recently by the MILAGRO collaboration \cite{milagro1} 
$F_{MILAGRO}\lsim (0.3\div 0.6)\, F_{\rm Crab}$.

\section{Discussion}

The model of particle acceleration near the horizon of a supermassive
black hole which we have discussed in this paper is based on a number
of assumptions: maximum rotation moment of the black hole, low matter
and radiation density in the acceleration volume, absence of back
reaction of the accelerated particles and their radiation on the EM
field, uniform magnetic field at large distance from the black
hole. These assumptions have one common feature: they facilitate
acceleration to higher energies and minimize losses (and, therefore,
the produced radiation). We have found that even in these idealized
conditions the acceleration of protons to energy $E= 10^{20}$~eV
requires extreme values of parameters, $M\simeq 10^{10} M_{\odot}$ and
$B\simeq 3\times 10^4$~G. Moreover, the acceleration is very
inefficient: the total power emitted in TeV gamma rays is $100-1000$
times larger than in UHECR. In view of recent TeV observations, this
rules out some UHECR models based on this mechanism of acceleration,
e.g. the model of several nearby dormant galactic nuclei (``dead
quasars'') which was aimed to explain observed UHECR flux with energy
$E>10^{20}$~eV.

In a more realistic case the above conditions may not be satisfied completely,
and the acceleration of protons to energy $E\simeq 10^{20}$~eV in the
continuous regime may not be possible. The synchrotron losses due to the
presence of random component $B_{\rm rand}$ of the magnetic field can be
neglected if
\begin{equation}
B_{\rm rand}\ll \frac{B}{\cal R}\,\frac{m}{m_p}\simeq 
10^{-2}\,\frac{m}{m_p}\, B\; ,
\end{equation}
where ${\cal R}$ is given by (\ref{eq:E_EM/E_UHECR}).
This means that the presence of a tiny (1\% level) random magnetic field
will lead to the decrease of maximal energies of accelerated protons 
and increase of electromagnetic luminosity of the accelerator. Note that the
synchrotron radiation is emitted in this case at energies
\begin{equation}
\label{synch1}
\epsilon_{\rm synch}\le \frac{m}{e^2}\,\frac{B}{B_{\rm rand}}
\simeq 0.1\,\frac{m}{m_p}\,\frac{B}{B_{\rm rand}} \mbox{ TeV}\; .  
\end{equation}
The power is still given by the
Eq. (\ref{eq:TeV_luminosity}). 

Even if the strength of the random component of magnetic field is as small as
$10^{-5}\;B$, for electrons, which are inevitably present in the accelerator,
the synchrotron losses will dominate over the curvature losses. The
electromagnetic power emitted by electrons will be in the 100~MeV - 10~TeV
energy band (see Eq. (\ref{synch1})). Assuming that the density of electrons
is of the same order as the density of protons, one obtains the same estimate
(\ref{eq:TeV_luminosity}) for the 100 MeV luminosity of the accelerator. This
means that such an accelerator is not only a powerful TeV source, but is also,
in fact, an extremely powerful EGRET source.

Even if the idealized conditions are realized in Nature, corresponding objects
must be extremely rare. Thus, only a very small fraction of (active or quiet)
galactic nuclei could be stationary sources of UHE protons with energies above
$10^{20}$~eV.

If the parameters of the model are not precisely tuned to their optimal
values, one expects the maximum energies of accelerated protons to be somewhat
below $10^{20}$~eV. It is therefore interesting to note that most of the
correlations of UHECR with the BL Lacertae objects come from the energy range
$(4-5)\times 10^{19}$~eV. The central engine of BL Lacs is thought to consist
of a supermassive black hole; it is possible that the acceleration mechanism
considered above is operating in these objects \cite{neronov}.  This mechanism
may also operate in the centers of other galaxies which may possess
(super)massive black holes, including our own Galaxy where it may be
responsible for the production of cosmic rays of energies up to $\sim
10^{18}$~eV \cite{Levinson:2000zz,longpaper}.

The constraints from TeV observations are different in this case.  First,
cosmic rays of lower energies propagate over cosmological distances, so the
UHECR flux is collected from a much larger volume, and the number of sources
may be larger. Correspondingly, the TeV luminosity of each source is
smaller. Second, TeV radiation attenuates substantially over several hundred
megaparsecs. Third, at $E<10^{20}$~eV the ratio $L_{\rm EM}/L_{UHECR}$ is
smaller. Consider e.g. the case of $O(100)$ sources located at $z\sim 0.1$
with typical maximal energy at accelerator $E\sim 5\times 10^{19}$ eV.
According to Fig.~\ref{fig:energy}, in this case the typical energy of
produced $\gamma$-rays is $\sim 4$ TeV. The flux of $\gamma$-rays in this
energy range is attenuated by a factor $10 - 100$, while according to
Eq.~(\ref{eq:E_EM/E_UHECR}), $L_{\rm EM}/L_{UHECR} \sim 50$.  Therefore, one
may expect $F_{\rm EM} \sim (0.01 - 0.1)\, F_{\rm Crab}$ for the TeV flux from
each of these sources. This is within the range of accessibility of modern
telescopes. For example, the TeV flux from the nearby ($z=0.047$) BL Lac 1ES
1959+650, which correlates with arrival directions of UHECR
\cite{EgretBllacs,Gorbunov:2004bs}, is at the level of $0.06F_{\rm Crab}$
during quiet phase and rises up to $2.9 F_{\rm Crab}$ during flares. Several
other BL Lacs which are confirmed TeV sources have fluxes $\approx 0.03F_{\rm
Crab}$, see e.g. Ref.~\cite{Aharonian:2004kd}.

This paper mainly concerns the stationary regime of acceleration when the
acceleration volume is not polluted by the creation of $e^+e^-$ pairs. To
ensure this condition we required that magnetic field does not exceed the
critical value (\ref{eq:discharge_cond}). If the magnetic field is larger, the
acceleration by the mechanism considered here can only happen during flares
which are interrupted by the creation of $e^+e^-$ plasma and neutralization of
the electric field as discussed in the end of
Sect.~\ref{sect:discharge}. Although we do not have a quantitative model of a
flare, some features of this regime and its consequences for UHECR production
can be understood qualitatively. Since there is no constraint on the magnetic
field in this regime, the maximum energies of accelerated protons may exceed
$10^{20}$~eV. However, the efficiency of the acceleration during flares must
be much lower than in the stationary case. First, as follows from
Fig.~\ref{fig:ratio}, the ratio $L_{\rm EM}/L_{\rm UHECR}$ is larger at large
$B$. Second, the dominant part of the EM radiation is produced by the created
electrons and positrons whose number density exceeds by far the number density
of protons. Thus, one expects that the ratio $L_{\rm EM}/L_{\rm UHECR}$ for
this sources is much larger than in Eq.~(\ref{eq:TeV_luminosity}).

An UHECR accelerators operating in the flaring regime would produce
approximately constant UHECR flux at the Earth. The reason is the time delay
of protons due to random deflections in the extragalactic magnetic
fields. This delay is of the order of $\sim 10^5 [\alpha/1^{\circ}]^2$~yr for
a source at 100~Mpc, where $\alpha$ is the deflection angle. Since the time
scale of flares (light crossing time) is of the order of day(s), the
variations of UHECR flux would average away. On the contrary, the TeV
radiation from such a source would be highly variable, with powerful ``TeV
bursts'' and {\em average} energy flux in TeV band exceeding that in UHECR by
a factor of $10^4$ or higher. Note that there exist tight constraints on
transient TeV sources: the energy flux of a TeV burst of duration $10^5$~s has
to be less than $10^{-10}\mbox{erg/cm$^2$s}\sim 10\, F_{\rm Crab}$
\cite{milagro,tibet}. As in the case of stationary accelerator, this
constraint excludes the possibility to explain observed UHECR flux by a few
nearby proton accelerators operating in the flaring regime.  The hypothesis of
several hundred remote sources is not constrained by TeV observations.

To summarize, the model of compact UHE proton accelerators which operate near
the horizons of supermassive black holes in galactic nuclei can explain only
the sub-GZK events. A large number (several hundred) of sources situated at
cosmological distances is required.  Production of UHECR in such sources may
be associated with the blazar-type activity, TeV $\gamma$-radiation being an
important signature of the model, testable by the existing $\gamma$-ray
telescopes.

\begin{acknowledgments}
We are grateful to F. Aharonian, M. Kachelriess, D. Semikoz, M. Shaposhnikov
and M. Teshima for many useful discussions during various stages of this
project.  The work of P.T. is supported in part by the Swiss Science
Foundation, grant 20-67958.02.
\end{acknowledgments}

\end{document}